\documentclass[letterpaper, 10 pt, conference]{ieeeconf}  
\usepackage[table]{xcolor}

\IEEEoverridecommandlockouts                              

\usepackage{amsthm}
\usepackage{amssymb}
\usepackage{xcolor,graphicx,epstopdf,caption,subcaption}

\usepackage{multirow} 
\usepackage{mathtools,amsmath,amsgen,amstext,amsbsy,amsopn,amsfonts}
\newtheorem{assumption}{Assumption}

\newtheorem{theorem}{Theorem}

\newcommand{\bgeq}{\begin{equation}}
\newcommand{\edeq}{\end{equation}}
\newcommand{\bgdm}{\begin{displaymath}}
\newcommand{\eddm}{\end{displaymath}}

\title{\LARGE \bf
Composite Distributed Learning and
Synchronization of Nonlinear Multi-Agent
Systems with Complete Uncertain Dynamics
\thanks{This work is supported in part by the National Science Foundation under Grant CMMI-1952862 and CMMI-2154901.}
}

\author{Emadodin Jandaghi, Dalton L. Stein, Adam Hoburg, Paolo Stegagno, Mingxi Zhou, Chengzhi Yuan
\thanks{E. Jandaghi,  D. L. Stein and C. Yuan are with the  Department of Mechanical, Industrial and Systems Engineering, Kingston, RI 02881, USA, 
        {\tt\small emadjandaghi@uri.edu; daltonstein98@uri.edu; cyuan@uri.edu}. A. Huborg and P. Stegagno are with the Department of Electrical, Computer and Biomedical Engineering, Kingston, RI 02881, USA, 
        {\tt\small ajhoburg@uri.edu; pstegagno@uri.edu}. M. Zhou is with the Graduate School of Oceanography, , Kingston, RI 02881, USA, 
        {\tt\small mzhou@uri.edu}}%
    }
    
\begin{document}

\maketitle
\thispagestyle{empty}
\pagestyle{empty}

\begin{abstract}

This paper addresses the problem of composite synchronization and learning control in a network of multi-agent robotic manipulator systems with heterogeneous nonlinear uncertainties under a leader-follower framework. A novel two-layer distributed adaptive learning control strategy is introduced, comprising a first-layer distributed cooperative estimator and a second-layer decentralized deterministic learning controller. The first layer is to facilitate each robotic agent's estimation of the leader's information. The second layer is responsible for both controlling individual robot agents to track desired reference trajectories and accurately identifying/learning their nonlinear uncertain dynamics. The proposed distributed learning control scheme represents an advancement in the existing literature due to its ability to manage robotic agents with completely uncertain dynamics including uncertain mass matrices. This allows the robotic control to be environment-independent which can be used in various settings, from underwater to space where identifying system dynamics parameters is challenging. The stability and parameter convergence of the closed-loop system are rigorously analyzed using the Lyapunov method. Numerical simulations validate the effectiveness of the proposed scheme. 

\end{abstract}

\section{INTRODUCTION}
Robotics has many applications from manufacturing to surgical procedures \cite{cui2012mutual,seenu2020review,jandaghi2023motion,ghafoori2024novel}. However, controlling robots in space and underwater is challenging due to unpredictable robot dynamics in such environments. Also, the increasing demand for high precision and operational complexity has shifted the focus towards cooperatively utilizing multiple standard robots. This is beneficial with improved efficiency, cost reduction, and redundancy \cite{yuan2017formation,hazon2008redundancy}. To this end, numerous studies have been done to formulate diverse decentralized control strategies for coordinating multiple robotic arms  \cite{huang2015adaptive}.

Despite extensive research, one critical outstanding challenge is the management of model uncertainties which can impair the performance of distributed control systems \cite{abdelatti2018cooperative}. Some studies have examined the role of uncertainties in robotic arm control \cite{liu2021adaptive}, however, these did not extend their findings to the synchronization of multiple robots. \cite{rodriguez2004mutual} explored a virtual leader-follower strategy for robot manipulators under uncertainties and disturbances. While a high-gain observer was used for velocity estimation, this approach risks exciting unmodeled high-frequency dynamics and magnifying measurement noise. Such issues make controller implementation challenging and necessitate meticulous parameter tuning. Another drawback in the existing literature is the assumption of homogeneous system dynamics across all robots, which is often not the case in real-world applications. Meanwhile, \cite{chung2009cooperative} ignored system uncertainties and assumed full model knowledge. Furthermore, while some research has considered non-identical robotic systems \cite{rodriguez2004mutual}, others \cite{wang2013flocking} mostly focused solely on achieving adaptive tracking control, without consideration for the convergence of controller parameters to their optimal states. 
Previous literature has not fully tackled the learning of system uncertainties  without certain assumptions. These include having a known Mass or Inertia matrix \cite{dong2019composite}, and using large gains to suppress errors caused by the Coriolis and Centripetal force matrix \cite{liu2021adaptive}. 
In contrast, our work uniquely identifies the nonlinear uncertain dynamics of each robot without making any assumptions on certainty or structure of each system.

In this study, we make a significant step forward compared to previous research. This framework is completely independent of any multi-agent system attributes, rendering this method universally applicable to any nonlinear Euler-Lagrange (EL) system dynamics. We tackle the challenges of achieving synchronization control and integrating learning capabilities into multi-robot systems with heterogeneous nonlinear uncertain dynamics without any knowledge of each robotic system's nonlinearities. 
Our control architecture employs a fixed directed graph communication with a virtual leader with linear dynamics. 
The control strategy is dual-layered: The first layer focuses on cooperative estimation, allowing agents to have inter-agent communication and share leader-estimated states and system matrices.
The second layer involves a decentralized adaptive learning controller to regulate each robot's state and pinpoint its unique dynamics using first-layer estimates. The second layer operates without any data sharing, allowing each local robot to implement its adaptive learning controller in a completely decentralized way.
Our adaptive learning control law uses precise function approximation with Radial Basis Function (RBF) Neural Networks (NN) for the identification of systems uncertain dynamics. This enhances control over robotics in space and underwater environments, where the system dynamics are often unpredictable. We confirm the efficacy of the approach through mathematically rigorous analysis and comprehensive simulation studies. 


\section{PRELIMINARIES AND PROBLEM STATEMENT}\label{sec:prelim}
\label{sec:Preliminaries_Problem_Statement}
\subsection{Notation and Graph Theory}
We denote the set of real numbers as \scalebox{0.9}{$\mathbb{R}$}. \scalebox{0.9}{$\mathbb{R}^{m \times n}$} is the set of real \scalebox{0.9}{$m \times n$} matrices, and \scalebox{0.9}{$\mathbb{R}^n$} is the set of real \scalebox{0.9}{$n \times 1$} vectors. 
\scalebox{0.9}{$A \otimes B$} signifies the Kronecker product of matrices $A$ and $B$. 
Given two integers $k_1$ and \scalebox{0.9}{$k_2$ with $k_1 < k_2$, $\mathbf{I}[k_1, k_2] = \{k_1, k_1 + 1, \dots, k_2\}$}. For a vector \scalebox{0.9}{$x \in \mathbb{R}^n$}, its norm is defined as \scalebox{0.9}{$|x| := (x^T x)^{1/2}$}. For a square matrix \scalebox{0.9}{$A$, $\lambda_i(A)$ denotes its $i$}-th eigenvalue, while \scalebox{0.9}{$\lambda_{\text{min}}(A)$} and \scalebox{0.9}{$\lambda_{\text{max}}(A)$} represent its maximum and minimum eigenvalues, respectively. A directed graph \scalebox{0.9}{$G = (V, E)$} comprises nodes in the set \scalebox{0.9}{$V = \{1, 2, \dots, N\}$} and edges in \scalebox{0.9}{$E \subseteq V \times V$}. An edge from node $i$ to node $j$ is represented as \scalebox{0.9}{$(i, j)$}, with $i$ as the parent node and $j$ as the child node. Node $i$ is also termed a neighbor of node $j$. $N_i$ is considered as the subset of $V$ consisting of the neighbors of node $i$. A sequence of edges in $G$, \scalebox{0.9}{$(i_1, i_2), (i_2, i_3), \dots, (i_k, i_{k+1})$}, is called a path from node $i_1$ to node $i_{k+1}$. Node $i_{k+1}$ is reachable from node $i_1$. A directed tree is a graph where each node, except for a root node, has exactly one parent. The root node is reachable from all other nodes. A directed graph $G$ contains a directed spanning tree if at least one node can reach all other nodes. The weighted adjacency matrix of $G$ is a non-negative matrix \scalebox{0.9}{$A = [a_{ij}] \in \mathbb{R}^{N \times N}$}, where $a_{ii} = 0$ and \scalebox{0.9}{$a_{ij} > 0 \implies (j, i) \in E$}. The Laplacian of $G$ is denoted as \scalebox{0.9}{$L = [l_{ij}] \in \mathbb{R}^{N \times N}$}, where \scalebox{0.9}{$l_{ii} = \sum_{j=1}^{N} a_{ij}$} and \scalebox{0.9}{$l_{ij} = -a_{ij}$ if $i \neq j$}. 
From \cite{ren2005consensus}, $L$ has one zero eigenvalue and remaining eigenvalues with positive real parts if and only if $G$ has a directed spanning tree.
\subsection{Radial Basis Function NNs} \label{1b}
The RBF Neural Networks (NN) can be described as \scalebox{0.9}{$f_{nn}(Z)=\sum_{i = 1}^{N}w_i s_i(Z)=W^{T}S(Z)$}, where \scalebox{0.9}{$Z\in\Omega_Z\subseteq\mathbb{R}^q$} and \scalebox{0.9}{$W=w_1,...,w_N^T\in\mathbb{R}^N$} as input and weight vectors respectively. \scalebox{0.9}{$N$} indicates the number of NN nodes, \scalebox{0.9}{$S(Z)=[s_1(||Z-\mu_i||),...,s_N(||Z - \mu_i||)]^T$} with \scalebox{0.9}{$s_i(\cdot)$}  is a RBF, and \scalebox{0.9}{$\mu_i(i = 1,...,N)$} is distinct points in the state space.
The Gaussian function \scalebox{0.9}{$s_i (||Z - \mu_i||)=\operatorname{exp}\left[-\frac{(Z-\mu_i)^T(Z-\mu_i)}{\eta_i^2}\right]$} is generally used for RBF, where \scalebox{0.9}{$\mu_i=[\mu_{i1},\mu_{i2},...,\mu_{iN}]^T$} is the center and \scalebox{0.9}{$\eta_i$} is the width of the receptive field. The Gaussian function categorized by localized radial basis function \scalebox{0.9}{$s$} in the sense that \scalebox{0.9}{$  s_i (||Z - \mu_i||)\rightarrow 0 $} as \scalebox{0.9}{$ ||Z||\rightarrow \infty $}.
It has been shown in \cite{park1991universal}, for any continuous function \scalebox{0.9}{$  f(Z):\Omega_Z \rightarrow R $ where $  \Omega_Z \subset R^p $} is a compact set, there exists an ideal constant weight vector \scalebox{0.9}{$W^*$}, such that for each \scalebox{0.9}{$  \epsilon^* > 0 $}, \scalebox{0.9}{$   f(Z) = W^{*T}S(Z) + \epsilon(Z)$}, $  \forall Z \in \Omega_Z $, where $\epsilon(Z) $ is the approximation error which can be made arbitrarily small given a sufficiently large node number N.  Moreover, according to  \cite{wang2018deterministic}, given any continuous recurrent trajectory \scalebox{0.9}{$Z(t) : [0,\infty) \rightarrow R^q$, $Z(t)$} remains in a bounded compact set \scalebox{0.9}{$\Omega_{Z} \subset \mathbb{R}^q$}, for RBF NN \scalebox{0.9}{$ W^{T}S(Z)$} with centers placed on a regular
lattice (large enough to cover compact set \scalebox{0.9}{$\Omega_{Z}$}), the regressor subvector  \scalebox{0.9}{$S_{\zeta}(Z)$} consisting of
RBFs with centers located in a small neighborhood of \( Z(t) \) is persistently exciting (PE). 

\subsection{Problem Statement}
We consider a multi-robot manipulator system of $N$ robots with heterogeneous uncertain nonlinear dynamics:
\begin{equation} 
    M_i(\theta_i)\ddot{\theta}_i + C_i(\theta_i,\dot{\theta}_i)\dot{\theta}_i + g_i(\theta_i) = \tau_i, \quad i \in \mathbf{I}[1,N],
    \label{eqn:Config_Space}
\end{equation}
where \scalebox{0.9}{$ \tau_i \in \mathbb{R}^{n}$} is the vector of input signals. The subscript $i$ denotes the $i^{\textrm{th}}$ robotic agent. For each \scalebox{0.9}{$i \in \mathbf{I}[1,N]$}, \scalebox{0.9}{$\theta_i = [\theta_{i1}, \theta_{i2}, ..., \theta_{in}]^T \in \mathbb{R}^{n}$} are the joint positions, \scalebox{0.9}{$\dot{\theta}_i, \ddot{\theta}_i$} are the joint velocities and accelerations, respectively. \scalebox{0.9}{$M \in \mathbb{S}_+^n$} is a positive definite mass matrix, \scalebox{0.9}{$C \in \mathbb{R}^{n \times n}$} is the matrix containing Coriolis and Centripetal forces, \scalebox{0.9}{$g \in \mathbb{R}^{n}$} is the gravity term. 
It is crucial to note that as opposed to our previous work \cite{dong2019composite}, we make no assumptions about the certainty of any system matrices ($M,C,$ and $g$) for control design.
We rewrite the system dynamics as:
\begin{equation} \small
    \begin{aligned}
        &\dot{x}_{i1} = x_{i2} \\
        &\dot{x}_{i2} = M_i^{-1}(x_{i1})[\tau_i - C_i(x_i)x_{i2} - g_i(x_{i1})]
        \label{systemdynamics}
    \end{aligned}
\end{equation}
where \( x_{i1} = \theta_i \), \( x_{i2} = \dot{\theta}_i \), and let \( x_i \) be a column vector of \( x_{i1} \) and \( x_{i2} \).
The virtual leader's dynamics is:
\begin{equation} \small
    \begin{aligned}
        \dot{\chi}_0 = A_0\chi_0
        \label{leaderdynamics}
    \end{aligned}
\end{equation}
Here, "0" marks the leader node. The leader state \( \chi_0 \) is a column vector of \( x_{01} \) and \( x_{02} \). \( A_0 \) is a constant matrix. 
We make two assumptions, which do not sacrifice generality:
\begin{assumption}\label{assump1}
  All eigenvalues of \(A_{0}\) are imaginary.
\end{assumption}
\begin{assumption}\label{assump2}
  The digraph \(G\) has a directed spanning tree rooted at node 0.
\end{assumption}
Assumption \ref{assump1} ensures all the states of the leader dynamics remain periodic and uniformly bounded. The Laplacian \( L \) of the graph can be divided as follows using Assumption \ref{assump2}:
\scalebox{0.7}{$
L = \begin{bmatrix} 
\sum_{j=1}^{N} a_{0j} & -[a_{01}, \ldots, a_{0N}] \\
-\Phi_{1N} & H
\end{bmatrix}.
$} 
Here, \( \Phi \) is a diagonal matrix and \( H \) has only positive real parts for its non-zero eigenvalues.
  Given the robotic systems \eqref{systemdynamics} and a leader \eqref{leaderdynamics}, with Assumptions~\ref{assump1}-\ref{assump2}, we aim to design such a controller that:  (1) \textit{Cooperative Synchronization:} All robots should synchronize to the leader, i.e., \scalebox{0.9}{$\lim_{{t \to \infty}} (x_{i1}(t) - x_{01}(t)) = 0 \quad \forall i \in \mathbf{I}[1,N]$}. (2) \textit{Decentralized Learning:} Each robot should learn its own dynamics via its local adaptive controller using RBF NNs.

To this end, a two-layer framework is proposed: The first layer uses a distributed observer to estimate the leader's state and system information; The second layer uses a decentralized controller for synchronization control and dynamics learning.
Only the first layer requires data sharing between nearby robots and the second layer just works on local data. 

\section{DISTRIBUTED ADAPTIVE CONTROL}
\label{sec:Adaptive}
\subsection{First Layer: Distributed Cooperative Estimator}
In a distributed control setting, the leader's information including \( \chi_0 \) and \( A_0 \), may not be available to all agents. This constraint leads us to create a distributed cooperative estimator, enabling each agent to estimate the leader's state and dynamics through inter-agent collaboration:
\begin{equation} \small
\dot{\hat\chi}_i(t)=A_i(t)\hat\chi_i(t)+\beta_{i1}\sum_{j=0}^{N}a_{ij}(\hat\chi_j(t)-\hat\chi_i(t)),\ \forall i \in I[1, N]\
\label{5}
\end{equation}
The observer states for each robot \( \hat\chi_i = [\hat{x}_{i1},\hat{x}_{i2}]^T\) are used to estimate the leader's state \( \chi_0 = [x_{01},x_{02}]^T \). \( \beta_{i1} \) are positive design constant numbers. The time-varying system parameters \( \hat{A}_i(t) \) are updated by the following for all \( i \):
\begin{equation} \small
\dot{\hat{A}}_i(t) = \beta_{i2} \sum_{j=0}^{N} a_{ij} (\hat{A}_j(t) - \hat{A}_i(t)), \quad \forall i \in I[1, N]
\label{6}
\end{equation}
\( \hat{A}_i \) are used to estimate the leader's system matrix \( A_0 \) and have dimensions \( n \times n \). The constants \( \beta_{i2} \) are all positive numbers. 
By defining \( \tilde{\chi}_i = \hat{\chi}_i - \chi_0 \) and \( \tilde{A}_i = \hat{A}_i - A_0 \), we obtain the error dynamics for agent \( i \):
\begin{equation} \small
\begin{aligned} 
\dot{\tilde{\chi}}_i(t) &= A_0\tilde{\chi}_i(t) + \tilde{A}_i(t)\tilde{\chi}_i(t) + \tilde{A}_i(t)\chi_0(t) \\
&+ \beta_{i1} \sum_{j=0}^{N} a_{ij} (\tilde{\chi}_j(t) - \tilde{\chi}_i(t)), \quad \forall i \in I[1, N] \\ 
\dot{\tilde{A}}_i(t) &= \beta_{i2} \sum_{j=0}^{N} a_{ij} - (\tilde{A}_j(t) - \tilde{A}_i(t)) , \quad \forall i \in I[1, N].
\end{aligned}    
\end{equation}
Define \( \tilde{\chi} = \text{col}\{\tilde{\chi}_1, \ldots, \tilde{\chi}_N\} \), \( \tilde{A} = \text{col}\{\tilde{A}_1, \ldots, \tilde{A}_N\} \), \( \tilde{A}_b = \text{diag}\{\tilde{A}_1, \ldots, \tilde{A}_N\} \), \( B_{\beta_1} = \text{diag}\{\beta_{11}, \ldots, \beta_{N1}\} \), and \( B_{\beta_2} = \text{diag}\{\beta_{12}, \ldots, \beta_{N2}\} \), we can further obtain the error dynamics for the entire network:
\begin{equation} \small
    \begin{aligned}
\dot{\tilde{\chi}}(t) &= ((I_N \otimes A_0 ) - B_{\beta_1}(H \otimes I_{2n})) \tilde{\chi}(t) \\
&+ \tilde{A}_b(t) \otimes \tilde{\chi}(t) + \tilde{A}_b(t)(1_N \otimes \chi_0(t)), \\
\dot{\tilde{A}}(t) &= -B_{\beta_2}(H \otimes I_n) \tilde{A}(t).
\label{errorsystem}
\end{aligned}
\end{equation}
\begin{theorem} \label{theorem1}
    Given the error system \eqref{errorsystem} and under Assumptions~\ref{assump1} and \ref{assump2}, for all \( i \in I[1, N] \) and any initial conditions \( \chi_0(0), \hat{\chi}_i(0), \hat{A}_i(0) \), we have \( \lim_{t \to \infty} \tilde{A}_i(t) = 0 \) and \( \lim_{t \to \infty} \tilde{\chi}_i(t) = 0 \) exponentially.
\end{theorem}
The proof of Theorem~\ref{theorem1} is similar to that of \cite{dong2019composite}, which will be omitted here to save some space.
\subsection{Second Layer: Decentralized Learning Control}
Given that the leader's state information \( \chi_0 \) is unavailable to all robotic agents, \( \hat{\chi}_i \) will serve as the tracking reference signal for control design of each local agent. The learning control objective is met when each agent's system states \( x_i \) converge to \( \hat{\chi}_i \) in a decentralized manner. For stability analysis, we first state the following property of the robotic system: 
     the matrix \scalebox{0.9}{$\dot{M}_i(\theta_{i}) - 2C_i(\theta_i,\dot{\theta}_i)$} is skew-symmetric which means \scalebox{0.9}{$\frac{1}{2} \dot{M}_i(\theta_{i}) = 2C_i(\theta_i,\dot{\theta}_i)$}.
%
To this end, consider the \(i^{th}\) robot with a reference signal in its controller as \(\dot{x}_{ri} \) and \( \ddot{x}_{ri} \), while let a filtered output signal \(r_i\) be:
\begin{equation} \small
    r_i = \dot{x}_{i1} - \dot{x}_{ri} =\dot{e}_i + \lambda_i e_i \quad \forall i \in I[1, N]
\label{errorref}
\end{equation}
where \( \lambda_i > 0 \) and \(e_i \in \mathbb{R}^n\) is the position tracking error: $e_i = x_{i1} - \hat{x}_{i1}$. 
\( \dot{x}_{ri2} \) is the second derivative of reference signal with:
\begin{equation} \small
    M_i(x_{i1})\ddot{x}_{ri} + C_i(x_i,\dot{x}_i)\dot{x}_{ri} + g_i(x_i) = H_i(\chi_i)
    \label{eqn:parametricEL}
\end{equation}
where \scalebox{0.9}{$\chi_i = col\{x,\dot{x},\dot{x}_r,\ddot{x}_r\}$}.
The unknown nonlinear function  \( H_i(\chi_i) \) is aimed to be approximated using the RBF NN. Specifically, according to Section \ref{1b}, there exist RBF NNs \( {W^*_i}^T S_i(\chi_i) \) such that
$H_i(\chi_i) = W_i^{*T}S_i(\chi_i) + \epsilon_i(\chi_i), \ \forall i \in I[1, N]$, 
with \( W^*_i \) as the ideal constant weights, and \( |\epsilon_i(\chi_i)| \leq \epsilon^*_i \) is the ideal approximation errors which can be made arbitrarily small given sufficiently large number of neurons. Assuming \( \hat{W}_i \) as the estimate of \( W^*_i \), we construct the decentralized learning control law as:
\begin{equation}
\tau_i = \hat{W}_i^T S_i(\chi_i) - K_i r_i, \quad \forall i \in I[1, N].
\label{eqn:Controller}
\end{equation}
where \( \hat{W}_i^T S_i(\chi_i) = [\hat{W}_{i1}^T S_{i1}(\chi_i), \ldots, \hat{W}_{in}^T S_{in}(\chi_i)]^T \) and  \( K_i \in \mathbb{R}^{n \times n} \). A robust self-adaptation law for online updating \( \hat{W}_i \) is constructed as:
\begin{equation} \small
\begin{aligned}
    \dot{\hat{W}}_{ij} = -\Gamma_i [ S_{ij}(\chi_i) r_{ij} + \sigma_i \hat{W}_{ij} ], \  \forall i \in I[1, N], \  \forall j \in I[1, n].
    \label{eqn:Adaptive_Law}
\end{aligned}
\end{equation}
where \( \Gamma_i \) and \( \sigma_i \) are positive constant numbers with \( \sigma_i \) being very small, \( r_{ij} \) represents the \( j^{th} \) element of \( r_i \). Substituting \eqref{eqn:parametricEL} and \eqref{eqn:Controller} into \eqref{errorref} for all \( i \in I[1, N] \) yields:
\begin{equation} \small
\begin{aligned}
\dot{r}_i = M_i^{-1}(x_i)(\tilde{W}_i^T S_i(\chi_i) - \epsilon_i(\chi) - K_i r_i - C_i(x_i,\dot{x}_i) r_i)
\label{eqn:Error_Dynamics}
\end{aligned}
\end{equation}
where \( \tilde{W}_i := \hat{W}_i - W^*_i \). 
\begin{theorem} \label{theorem2}
    Given systems \eqref{eqn:Adaptive_Law} and \eqref{eqn:Error_Dynamics}. If there exists a sufficiently large compact set \scalebox{0.9}{$\Omega_{\chi_i}$} such that \scalebox{0.9}{$\chi_i \in \Omega_{\chi_i} \ \forall i \in \mathbf{I}[1,N]$}, then for any bounded initial conditions with \scalebox{0.9}{$\hat{W}_i(0) = 0 (\forall i \in \mathbf{I}[1,N])$} we have: (i) all the signals in the system remain uniformly bounded; (ii) the position tracking error \scalebox{0.9}{$x_{i1} - \hat{x}_{i1}$} converges exponentially to a small neighborhood around the origin, by choosing the design parameters with \scalebox{0.9}{$K_i \in \mathbb{S}_+^n \ \forall i \in \mathbf{I}[1,N]$}. (iii) along the system trajectory denoted by \scalebox{0.9}{$\varphi(\chi_i(t))|_{t \geq T_i}$} starting from $T_i$ which represents the settling time of tracking control, the local estimated neural weights \scalebox{0.9}{$\hat{W}_{i\psi}$} converge to small neighborhoods close to the corresponding ideal values \scalebox{0.9}{$W_{i\phi}^*$}, and locally-accurate identification of nonlinear uncertain dynamics defined in \eqref{eqn:parametricEL} can be obtained by \scalebox{0.9}{$\hat{W}^{T}_iS_i(\chi_i)$} as well as \scalebox{0.9}{$\bar{W}_i^TS_i(\chi_i)$} along the system trajectory \scalebox{0.9}{$\psi(\chi_i(t))|_{t \geq T_i}$}, where 
    \begin{equation} \small
    \begin{aligned}
        \bar{W}_i^T = mean_{t \in [t_{ia},t_{ib}]} \hat{W}_i(t), \quad \forall i \in \mathbf{I}[1,N],
    \end{aligned}
    \end{equation}
    with \scalebox{0.9}{$[t_{ia}, t_{ib}] (t_{ib}>t_{ia}>T_i)$} being a time segment after the transient period of tracking control. 
\end{theorem}
\textbf{Proof.} 
(i) Consider the following Lyaponuv function candidate for \eqref{eqn:Adaptive_Law} and \eqref{eqn:Error_Dynamics}:
\begin{equation} \small
    V_i = \frac{1}{2}r_i^TM_i(x_i)r_i + \frac{1}{2}\sum_{j=1}^n\tilde{W}_{ij}^{T} \Gamma^{-1} \tilde{W}_{ij}.
    \label{eqn:Lyaponuv_Fucntion_Candidate}
\end{equation}
Temporal differentiation of \eqref{eqn:Lyaponuv_Fucntion_Candidate} yields,
\begin{equation} \small
    \dot{V}_i = r_i^TM_i(x)\dot{r}_i + \frac{1}{2}r_i^T\dot{M}_i(x)r_i + \sum_{j=1}^n\dot{\tilde{W}}_{ij}^{T} \Gamma_i^{-1} \tilde{W}_{ij}. 
    \label{eqn:Time_Derivative_init}
\end{equation}
Exploiting \eqref{eqn:Error_Dynamics} yields
\begin{equation} \small
    \begin{split}
    \begin{aligned}
        \dot{V}_i = -r_i^TK_ir_i - r_i^T\epsilon(\chi_i) + r_i^T\tilde{W}_i^TS_i(\chi) + \sum_{j=1}^n\dot{\tilde{W}}_{ij}^{T} \Gamma_i^{-1} \tilde{W}_{ij}.
        \end{aligned}
    \end{split}
    \label{eqn:Proof_init}
\end{equation}
Utilizing \scalebox{0.9}{$\dot{\tilde{W}} = \dot{\hat{W}}$} in \eqref{eqn:Adaptive_Law}, and substituting \eqref{eqn:Adaptive_Law} into \eqref{eqn:Proof_init},
\begin{equation} \small
    \begin{split}
    \begin{aligned}
        \dot{V}_i &= -r_i^TK_ir_i - r_i^T\epsilon(\chi_i) +  r_i^T\tilde{W}_i^TS_i(\chi_i)  \\
        &+ \sum_{j = 1}^{n}\left(-\Gamma_i[S_{ij}(\chi_i)r_{ij} + \sigma_i\hat{W}_{ij}]\right)^T\Gamma_i^{-1}\tilde{W}_{ij} \\
        & = -r_i^TK_ir_i - r_i^T\epsilon(\chi_i) - \sum_{j=1}^n\sigma_i \hat{W}_{ij}^T \tilde{W}_{ij}
    \end{aligned}
    \end{split}
\end{equation}
Select \scalebox{0.9}{\( K_i = K_{i1} + K_{i2} \)} with \( K_{i1} \) and \( K_{i2} \) being positive definite to yield:
\begin{equation} \small
    \begin{aligned}
        \dot{V}_i = -r_i^T K_{i1} r_i - r_i^T K_{i2} r_i - r_i^T\epsilon (\chi_i) - \sum_{j=1}^{n} \sigma_i \hat{W}_i^T j \tilde{W}_{ij}.
    \end{aligned} 
\end{equation}
As a result, we have for all \scalebox{0.9}{$i \in \mathbf{I}[1,N]$} and \scalebox{0.9}{$j \in \mathbf{I}[1,n]$}:
\begin{equation} \small
    \begin{aligned}
        -\sigma_i \hat{W}_i^T j \tilde{W}_{ij} &= -\sigma_i (\tilde{W}_{ij} + W^*_{ij})^T \tilde{W}_{ij}  
        \leq -\sigma_i \tilde{W}_{ij}^2 - \sigma_i \tilde{W}_{ij} W^*_{ij}  \\
        &\leq -\frac{\sigma_i}{2} \|\tilde{W}_{ij}\|^2 + \frac{\sigma_i}{2} \|W^*_{ij}\|^2 .
    \end{aligned}
    \label{eqn:equality_1}
\end{equation}
Using the same approach we have:
\begin{equation} \small
    -r_i^TK_{i2}r_i - r_i^T\epsilon_i \leq \frac{\epsilon_i^T\epsilon_i}{4\lambda_{\text{min}}(K_{i2})}\leq \frac{||\epsilon_i^*||^{2}}{4 \lambda_{\text{min}}(K_{i2})} .
    \label{eqn:equality_2}
\end{equation}
Based on \eqref{eqn:equality_1} and \eqref{eqn:equality_2}, we conclude:
\begin{equation} \small
    \begin{aligned}
    \dot{V}_i \leq &-r_i^T K_{i1} r_i + \frac{||\epsilon_i^*||^2}{4 \lambda_{\text{min}}(K_{i2})} \\ &- \frac{1}{2} \sum_{j=1}^{n} \sigma_i \| \tilde{W}_{ij} \|^{2}  
    +\frac{1}{2} \sum_{j=1}^{n} \sigma_i \| W^*_{ij} \|^{2}. \quad \forall i \in I[1, N]  
    \end{aligned}
    \label{eqn:Lyaponuv_Final}
\end{equation}
Given that \( K_{i1} \) is positive definite, we can conclude that \( \dot{V}_i \) is negative definite under the condition that:
\begin{equation} \small
    \begin{aligned}
        \| r_i \|  > \frac{\| \epsilon^*_i \|}{\sqrt{2 \lambda_{\text{min}}(K_{i1}) \lambda_{\text{min}} (K_{i2})}} + \sqrt{\frac{2 \sigma_i}{\lambda_{\text{min}}(K_{i1})}} \sum_{j=1}^{n} \| W^*_{ij} \|
    \end{aligned}
\end{equation}
\begin{equation} \small
    \begin{aligned}
 \text{or} \quad \sum_{j=1}^{n} \| \tilde{W}_{ij} \| & > \frac{\| \epsilon_i \|}{\sqrt{2 \sigma_i \lambda_{\text{min}}(K_{i2})}} + \sum_{j=1}^{n} \| W^*_{ij} \|. \quad \forall i \in I[1, N] \nonumber
    \end{aligned}
\end{equation}
The signals \( r_i \) and \scalebox{0.9}{\( \tilde{W}_{ij} \)} are bounded, leading to boundedness of \scalebox{0.9}{\( \tilde{W}_{ij} \)}. As a result, \scalebox{0.9}{\( \hat{W}_{ij} \)} is also bounded. Given that the regressor vector \scalebox{0.9}{\( S_{ij}(\chi_i) \)} is bounded according to \cite{wang2009deterministic}, the feedback control law \( \tau_i \) from \eqref{eqn:Controller} is bounded as well. 

(ii) For the second part of the proof, we examine the Lyapunov function for the dynamics of $r_i$ given by \eqref{eqn:Error_Dynamics} as,
\begin{equation} \small
    V_{ri} = \frac{1}{2} r_i^T M_i(x_{i1}) r_i.
    \label{eqn:Second_Lyaponuv}
\end{equation}
Temporal differentiation of \eqref{eqn:Second_Lyaponuv} and substituting in \eqref{eqn:Error_Dynamics} for \scalebox{0.9}{$M_i(\theta)\dot{r}_i$} yields \scalebox{0.9}{\(\dot{V}_{ri} = -r_i^T K_i r_i + r_i^T \tilde{W}_i^T S_i(\chi_i)-\epsilon_i^T r_i\)}. 
Using an approach similar to the one used for inequality \eqref{eqn:equality_2}, and let \scalebox{0.9}{\( K_i = K_{i1} + 2K_{i2} \)}, we can demonstrate:
\begin{equation} \small
  \begin{aligned}
    -r_i^T K_{i2} r_i - r_i^T \epsilon_i \leq \frac{\epsilon_i^2}{4 \lambda_{\min}(K_{i2})} \leq \frac{\epsilon_i^{*2}}{4 \lambda_{\min}(K_{i2})}, \quad \forall i \in I[1, N]
\end{aligned}  
\end{equation}
\begin{equation} \small
 \begin{aligned}
    -r_i^T K_{i2} r_i + r_i \tilde{W}_i^T S_i(\chi_i)  \leq \frac{\tilde{W}_{i}^{*2} s_{i}^{*2}}{4 \lambda_{\min}(K_{i2})}, \quad \forall i \in I[1, N].
\end{aligned}   
\end{equation}
where \scalebox{0.9}{\( || S_i(\chi_i) || \leq s_i^* \)} for all \scalebox{0.9}{\( i \in I[1, N] \)}. The existence of such a \( s_i^* \) is confirmed by \cite{wang2009deterministic}. Therefore, we arrive at:
\begin{equation} \small
\begin{aligned}
  \dot{V}_{ri} \leq -r_i^T \lambda_{\min}(K_{i1}) r_i + \delta_i  \leq -\rho_i V_{ri} + \delta_i, \ \forall i \in I[1, N]  
  \label{31}
\end{aligned}    
\end{equation}
where $\rho_i = \min\{2 \lambda_{\min}(K_{i1}), \frac{2 \lambda_{\min}(K_{i1})}{\lambda_{\max}(M_i)} \}, \ 
\delta_i = \frac{\delta_i^{*2}}{4 \lambda_{\min}(K_{i2})} + \frac{\tilde{W}_{i}^{*2} s_{i}^{*2}}{4 \lambda_{\min}(K_{i2})}.$
Solving (\ref{31}) gives us:
\begin{equation} \small
  \begin{aligned}
  0 \leq V_{ri}(t) \leq V_{ri}(0) \exp(-\rho_i t) +\frac{\delta_i}{\rho_i}, \ \forall t \geq 0, \forall i \in I[1, N]  
\end{aligned}  
\end{equation}
This implies that there exists a finite time \scalebox{0.9}{\( T_i > 0 \)} such that \( r_i \) will exponentially converge to a small area around zero. This further confirms that the tracking errors \( e_i \) will also reach a small zone around zero. The size of this zone can be minimized by carefully choosing \( K_{i1} \) with \scalebox{0.9}{\( \lambda_{\min}(K_{i1}) > 0 \)}. 

(iii) From the above proof, since  \scalebox{0.9}{\( e_i = x_{i1} - \hat{x}_{i1} \)}, \scalebox{0.9}{\( \hat{x}_{i1} \)} will eventually converge to \scalebox{0.9}{\( x_{01} \)}, and \scalebox{0.9}{\( x_{01} \)} is a periodic signal under Assumption 1, \scalebox{0.9}{\( x_{i1} \)} will also become a periodic signal after a finite time \scalebox{0.9}{\( T_i \)}. Furthermore, while \scalebox{0.9}{\( e_i \)} converges to zero, \scalebox{0.9}{\( \dot{e_i} \)} converges to zero, leading \scalebox{0.9}{\( x_{i2} \)} converges to \scalebox{0.9}{\( {x}_{02} \)}. Since the leader dynamics is a smooth continuous LTI system, periodicity of \scalebox{0.9}{\( x_{01} \)} implies that \scalebox{0.9}{\( \dot{x}_{01} \)} is also periodic and thus \scalebox{0.9}{\( x_{i2} \)} is periodic after a finite time \scalebox{0.9}{\( T_i \)}.
Consequently, the inputs of RBF NNs \scalebox{0.9}{\( (\chi_i) \)} are made as periodic signals for all \scalebox{0.9}{\( t \geq T_i \)}. According to Section \ref{1b}, the partial PE condition of the localized RBF NN regression subvector \scalebox{0.9}{\( S_{i\phi}(\chi_i) \)} along the system trajectory \scalebox{0.9}{\( \phi_i(\chi_i(t)|_{t \geq T_i} \)} is guaranteed, where \scalebox{0.9}{\( \hat{W}_{i}^T S_{i}(\chi_i) = \hat{W}_{i\phi}^T S_{i\phi}(\chi_i) +\hat{W}_{i\phi^{\bar{}}}^T S_{i\phi^{\bar{}}}(\chi_i) \)}.
Accordingly, system dynamics of (19) can be expressed as:
\scalebox{0.9}{\( \dot{r}_i = M_i^{-1}(x_{i1})(\widetilde{W}_{i\phi}^T S_{i\phi}(\chi_i) - K_i r_i - C_i(\chi_i) r_i - \epsilon'_{i\phi}) \)}, 
where \scalebox{0.9}{\( \epsilon'_{i\phi} = \epsilon_i - \hat{W}_{i\phi^{\bar{}}}^T S_{i\phi^{\bar{}}}(\chi_i)\)} is the localized ideal NN approximation error along the tracking trajectory. Thus, the overall closed-loop adaptive learning system can be described by:
\begin{equation}
    \begin{aligned}
    \scalebox{0.63}{$
        \begin{pmatrix}
\dot{r}_i \nonumber \\
\dot{\widetilde{W}}_{i\phi, 1} \\
\dot{\widetilde{W}}_{i\phi, 2} \\
\vdots \\
\dot{\widetilde{W}}_{i\phi, n}
\end{pmatrix}
=
\begin{pmatrix}
-M_i^{-1}(x_{i1}) K_i & 
 M_i^{-1}(x_{i1}) \begin{bmatrix}
S_{i\phi, 1}^T(\chi_i) & 0 & 0 & 0 \\
0 & S_{i\phi, 2}^T(\chi_i) & 0 & 0 \\
\vdots & \vdots & \ddots & \vdots \\
0 & 0 & 0 & S_{i\phi, n}^T(\chi_i)
\end{bmatrix} \\
-\Gamma_i S_{i\phi, 1}(\chi_i) \\
-\Gamma_i S_{i\phi, 2}(\chi_i) \\
\vdots  & 0\\
-\Gamma_i S_{i\phi, n}(\chi_i)
\end{pmatrix}
$}
    \end{aligned}
\end{equation}
\begin{equation}
    \begin{aligned}
    \scalebox{0.63}{$
    \times \begin{pmatrix}
r_i \\
\widetilde{W}_{i\phi, 1} \\
\widetilde{W}_{i\phi, 2} \\
\vdots \\
\widetilde{W}_{i\phi, n}
\end{pmatrix}
+
\begin{pmatrix}
-M_i^{-1}(x_{i1}) \epsilon_{i\phi} \\
-\Gamma_i \sigma_i \widehat{W}_{i\phi, 1} \\
-\Gamma_i \sigma_i \widehat{W}_{i\phi, 2} \\
\vdots \\
-\Gamma_i \sigma_i \widehat{W}_{i\phi, n} 
\end{pmatrix}
$}
, \forall i \in I[1, N]
    \end{aligned}
\end{equation}

\begin{equation}
    \begin{aligned}
    \scalebox{0.63}{$
\begin{pmatrix}
\dot{\widetilde{W}}_{i\phi\bar{}, 1} \\
\dot{\widetilde{W}}_{i\phi\bar{}, 2} \\
\vdots \\
\dot{\widetilde{W}}_{i\phi\bar{}, n}
\end{pmatrix}
=
\begin{pmatrix}
-\Gamma_i (S_{i\phi\bar{}, 1} (\chi_i) r_{i} + \sigma_i \hat{W}_{i\phi\bar{}, 1}) \\
-\Gamma_i (S_{i\phi\bar{}, 2} (\chi_i) r_{i} + \sigma_i \hat{W}_{i\phi\bar{}, 2}) \\
\vdots \\
-\Gamma_i (S_{i\phi\bar{}, n} (\chi_i) r_{i} + \sigma_i \hat{W}_{i\phi\bar{}, n})
\end{pmatrix}
$}
, \forall i \in I[1, N]
    \end{aligned}
\end{equation}
Based on \cite{wang2009deterministic}, the local approximation error \( \epsilon'_{i\phi} \) and \scalebox{0.9}{\( \hat{W}_{i\phi^{\bar{}}}^T S_{i\phi^{\bar{}}}(\chi_i) \)} are both small. Moreover, \( \epsilon'_{i\phi} \) is proportional to \( \epsilon_i \). Studies \cite{wang2009deterministic} \cite{yuan2011persistency} have thoroughly examined the stability and convergence of the above closed-loop system. Specifically, it's established that the PE condition of \( S_{i\phi}(\chi_i) \) leads to exponential convergence of \( (r_i, \tilde{W}_{i\phi}) \) to zero. 

Further, since \scalebox{0.9}{\(\epsilon_{i\phi}\)} is proportional to \scalebox{0.9}{\( \epsilon_i \)}, and \scalebox{0.9}{\( \sigma_i \)} can be made as small as desired, \scalebox{0.9}{\( \tilde{W}_{i\phi} \)} will also converge to an arbitrarily small vicinity of zero. As such, we have $H_i(\chi_i) = \hat{W}_{i\phi}^T S_{i\phi}(\chi_i) + \epsilon_{i\phi, 1} = \bar{W}_{i\phi}^T S_{i\phi}(\chi_i) + \epsilon_{i\phi, 2}$, $\forall i \in I[1, N]$, where \( \epsilon_{i\phi, 1} \) and \( \epsilon_{i\phi, 2} \) are approximation errors. They are proportional to \( \epsilon_{i\phi} \) due to the proven convergence of \( \tilde{W}_{i\phi} \) to zero.
As for neurons far from the trajectory, \( S_{i\phi^{\bar{}}}(\chi_i) \) is minimal, impacting the neural weight adaptation only slightly. Therefore, the full RBF NN can still accurately approximate the unknown function \( H_i(\chi_i) \) along the trajectory for \( t \geq T_i \). I.e., $Hi(\chi_i) = \hat{W}_{i}^T S_{i}(\chi_i) + \epsilon_{i1} = \bar{W}_{i}^T S_{i}(\chi_i) + \epsilon_{i2}$. 
The approximation errors \( \epsilon_{i1} \) and \( \epsilon_{i2} \) are proportional to \( \epsilon_{i\phi, 1} \) and \( \epsilon_{i\phi, 2} \) respectively. This concludes the proof.

\section{SIMULATION}
\label{simulation}
The multiple 2-DOF robot manipulator system, as described in \eqref{systemdynamics}, is considered with parameters:
{\small
\begin{align}
\begin{aligned}
M_i(q_i) &= \begin{bmatrix} M_{i11} & M_{i12} \\ M_{i21} & M_{i22} \end{bmatrix}, 
C_i(q_i, \dot{q}_i) = \begin{bmatrix} C_{i11} & C_{i12} \\ C_{i21} & C_{i22} \end{bmatrix}, \\ \nonumber
F_i(\dot{q}_i) &= \begin{bmatrix} F_{i11} \\ F_{i21} \end{bmatrix}, 
g_i(q_i) = \begin{bmatrix} g_{i11} \\ g_{i21} \end{bmatrix}.\\
M_{i11} &= m_{i1}l_{ic1}^2 + m_{i2}(l_{i1}^2 + l_{ic2}^2 + 2l_{i1}l_{ic2}\cos(qi2))  + I_{i1} + I_{i2}, \\
M_{i12} &= m_{i2}(l_{ic2}^2 + l_{i1}l_{ic2}\cos(q_{i2})) + I_{i2}, \\
M_{i21} &= m_{i2}(l_{ic2} + l_{i1}l_{ic2}\cos(qi2)) + I_{i2}, \ 
M_{i22} = m_{i2}l_{ic2}^2 + I_{i2}, \\
C_{i11} &= -m_{i2}l_{i1}l_{ic2}\dot{q}_{i2}\sin(q_{i2}), \ G_{i22} = m_{i2}l_{ic2}g\cos(q_{i1} + q_{i2}) \\
C_{i12} &= -m_{i2}l_{i1}l_{ic2}(\dot{q}_{i1} + \dot{q}_{i2})\sin(q_{i2}), \\
C_{i21} &= m_{i2}l_{i1}l_{ic2}\dot{q}_{i1}\sin(q_{i2}), \
C_{i22} = 0, \\
g_{i11} &= (m_{i1}l_{ic2} + m_{i2}l_{i1})g\cos(q_{i1})  + m_{i2}l_{ic2}g\cos(q_{i1} + q_{i2}).
\end{aligned}
\end{align}
}
%
%

where \scalebox{0.9}{\( l_{ic1}, l_{ic2} \)} represent half of link lengths \scalebox{0.9}{\( l_{i1}, l_{i2} \)}, respectively, and \scalebox{0.9}{\( F_{i11}, F_{i21} \)} are defined as constants. The inertia of the links is given by \scalebox{0.9}{\( I_{i1} \)} and \scalebox{0.9}{\( I_{i2} \)}, and the detailed values are provided in Table~\ref{tab:robot-parameters}. By employing \scalebox{0.9}{\( N = 5 \)} manipulators, the robotic agents are set to follow the reference trajectory provided by the virtual leader. A leader dynamics is formulated to generate a periodic signal for the synchronization control:
{\small
\[
\begin{bmatrix}
\dot{x}_{01} \\
\dot{x}_{02}
\end{bmatrix}, = \begin{bmatrix}
0 & 0 & 1 & 0 \\
0 & 0 & 0 & 1 \\
-1 & 0 & 0 & 0 \\
0 & -1 & 0 & 0
\end{bmatrix}
\begin{bmatrix}
x_{01} \\
x_{02}
\end{bmatrix},
\begin{bmatrix}
x_{01}(0) \\
x_{02}(0)
\end{bmatrix} = \begin{bmatrix}
0 \\
0.8 \\
0.8 \\
0
\end{bmatrix}
\]
}
\begin{figure}
\setlength{\belowcaptionskip}{-15pt} 
\centerline{\includegraphics[scale=0.06]{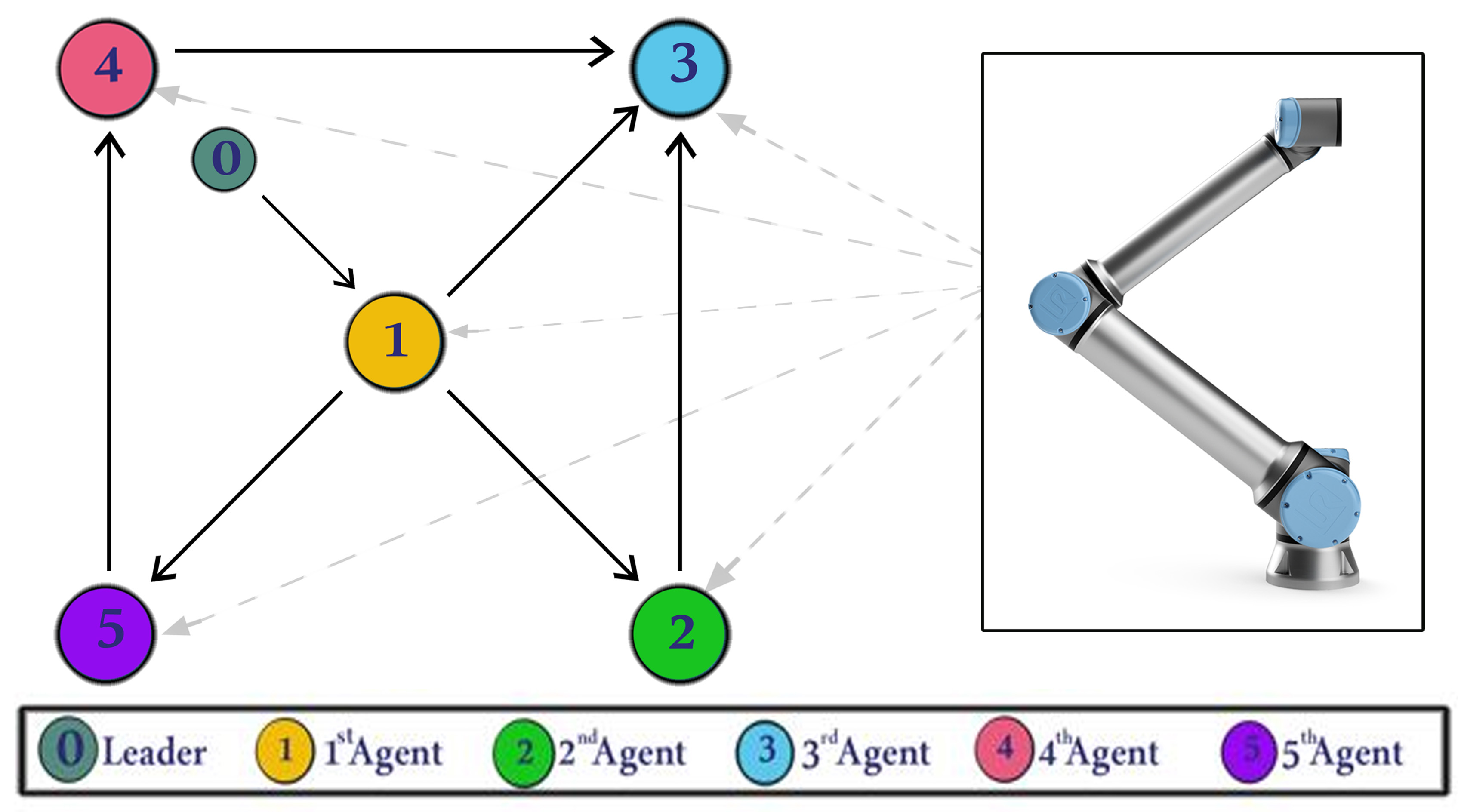}}
\caption{Network Topology with Agent 0 as the Virtual Leader}
\label{topology}
\end{figure}

\begin{figure}[!t]
\setlength{\belowcaptionskip}{-10pt} 
\centerline{\includegraphics[width=\columnwidth]{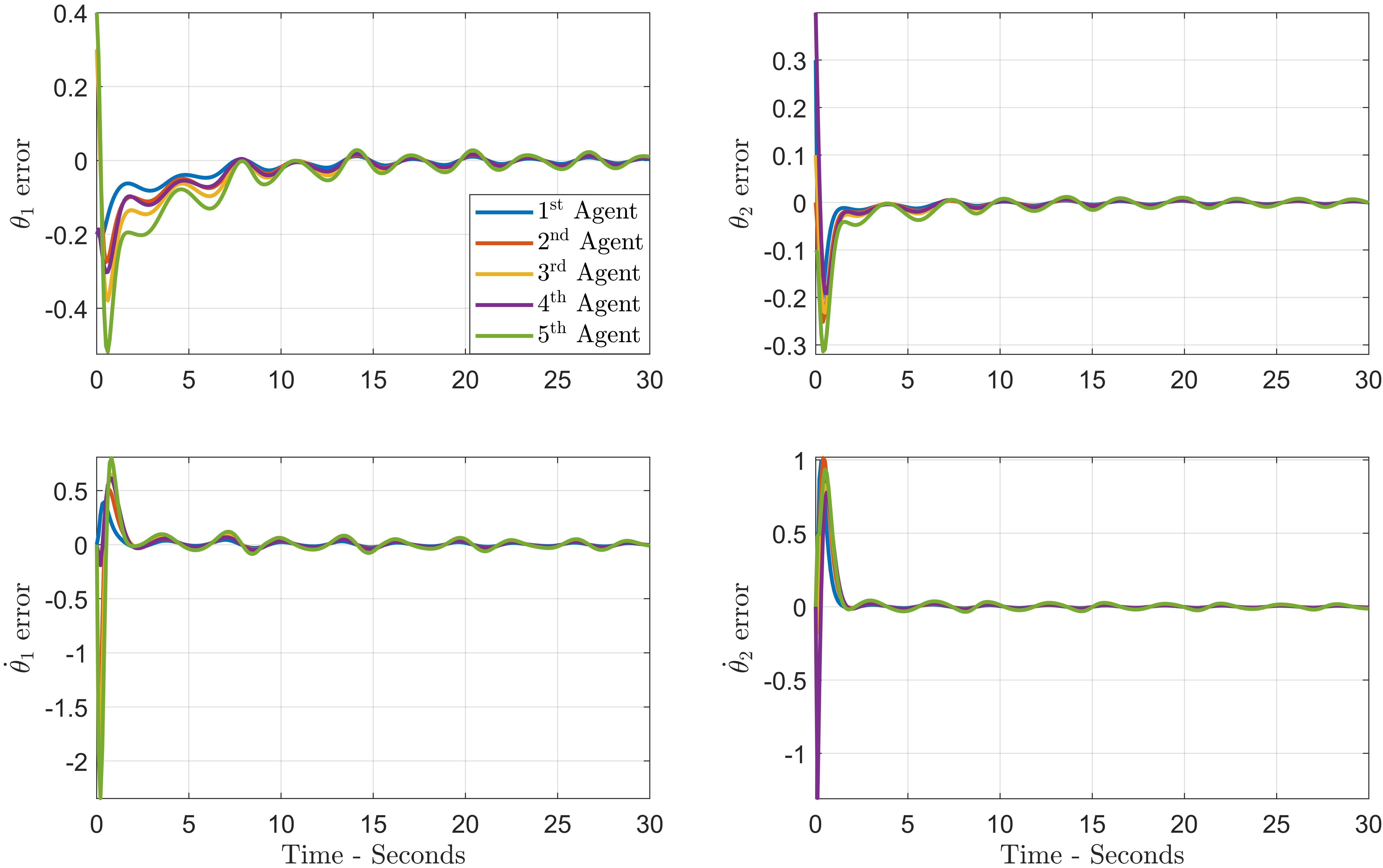}}
\caption{Tracking error for each agent.}
\label{fig3}
\end{figure}

The network topology \( G \) amongst the five robots is depicted in Fig~\ref{topology}. 
The system states encompass both measured signals and reference signals, totaling eight signals. If we use a neural network with 4 nodes for each signal, we end up with a massive number of nodes $4^8=65,536$ which is computationally expensive to train. However, the reference signals are expressed in terms of the measured signals, as shown in \eqref{errorref}, which accentuates the significance of the measured signals. Consequently, we construct the RBF NN using the dominant four dimensions of the system state. This strategy reduces the size of the NN to $4^4 = 256$ which not only saves computational resources but also allows us to build a more precise model by dedicating more nodes to each of these four key dimensions.
The range of each dimension lies within $[-1.2, 1.2]$, with a width parameter \scalebox{0.9}{$\gamma_i = 0.8$}. The observer and controller parameters are set to \scalebox{0.9}{$\beta_1 = \beta_2 = 1$, $\gamma_i = 10$}, \scalebox{0.9}{$K = 10$}, and \scalebox{0.9}{$\sigma_i = 0.001$} for all \scalebox{0.9}{$i \in [1, 5]$}. The initial conditions are defined as \scalebox{0.9}{$x_{11}(0) = [0.2\ 0.1]^T$}, \scalebox{0.9}{$x_{21}(0) = [0.3\ 0.5]^T$}, \scalebox{0.9}{$x_{31}(0) = [0.4\ 0.1]^T$, $x_{41}(0) = [0.2\ 0.5]^T$}, \scalebox{0.9}{$x_{51}(0) = [0.4\ 0.1]^T$}, and \scalebox{0.9}{$x_{i2} = [0\ 0]^T$, $\forall i \in [1, 5]$}. Initial conditions for all distributed observer states \scalebox{0.9}{$\hat{\chi}_i$} and NN weights \scalebox{0.9}{$\hat{W}_i$} are uniformly initialized to zero.
We conducted an assessment of two key components: the distributed cooperative estimator \eqref{5} and \eqref{6}, and the decentralized learning control law \eqref{eqn:Controller} and \eqref{eqn:Adaptive_Law}. Despite varying complexities and nonlinear uncertainties among multiple robotic agents, the results indicate satisfactory tracking performance, as confirmed by Fig. \ref{fig3}, which shows a rapid convergence of tracking errors to zero.
{\scriptsize 
\begin{table}
\centering
\caption{Parameters of the robot.}
\begin{tabular}{lcccccc}
\hline
Parameter & \multicolumn{6}{c}{Robot number} \\
\cline{2-7}
 & & 1 & 2 & 3 & 4 & 5 \\
\hline
\(m1\) (kg) & & 2 & 2.2 & 2.3 & 1.9 & 2.4 \\
\(m2\) (kg) & & 0.85 & 0.9 & 1 & 0.9 & 1.5 \\
\(l1\) (m) & & 0.35 & 0.5 & 0.6 & 0.52 & 0.57 \\
\(l2\) (m) & & 0.31 & 0.4 & 0.5 & 0.48 & 0.53 \\
\(I1 \times 10^{-3}\) (kg\,m\(^2\)) & & 61.25 & 70 & 72.14 & 67.21 & 73.42 \\
\(I2 \times 10^{-3}\) (kg\,m\(^2\)) & & 20.42 & 25.21 & 27.1 & 25.4 & 22.63 \\
\hline
\end{tabular}
\label{tab:robot-parameters}
\end{table}
}
\begin{figure}[!t]
\setlength{\belowcaptionskip}{-10pt} 
\centerline{\includegraphics[width=\columnwidth]{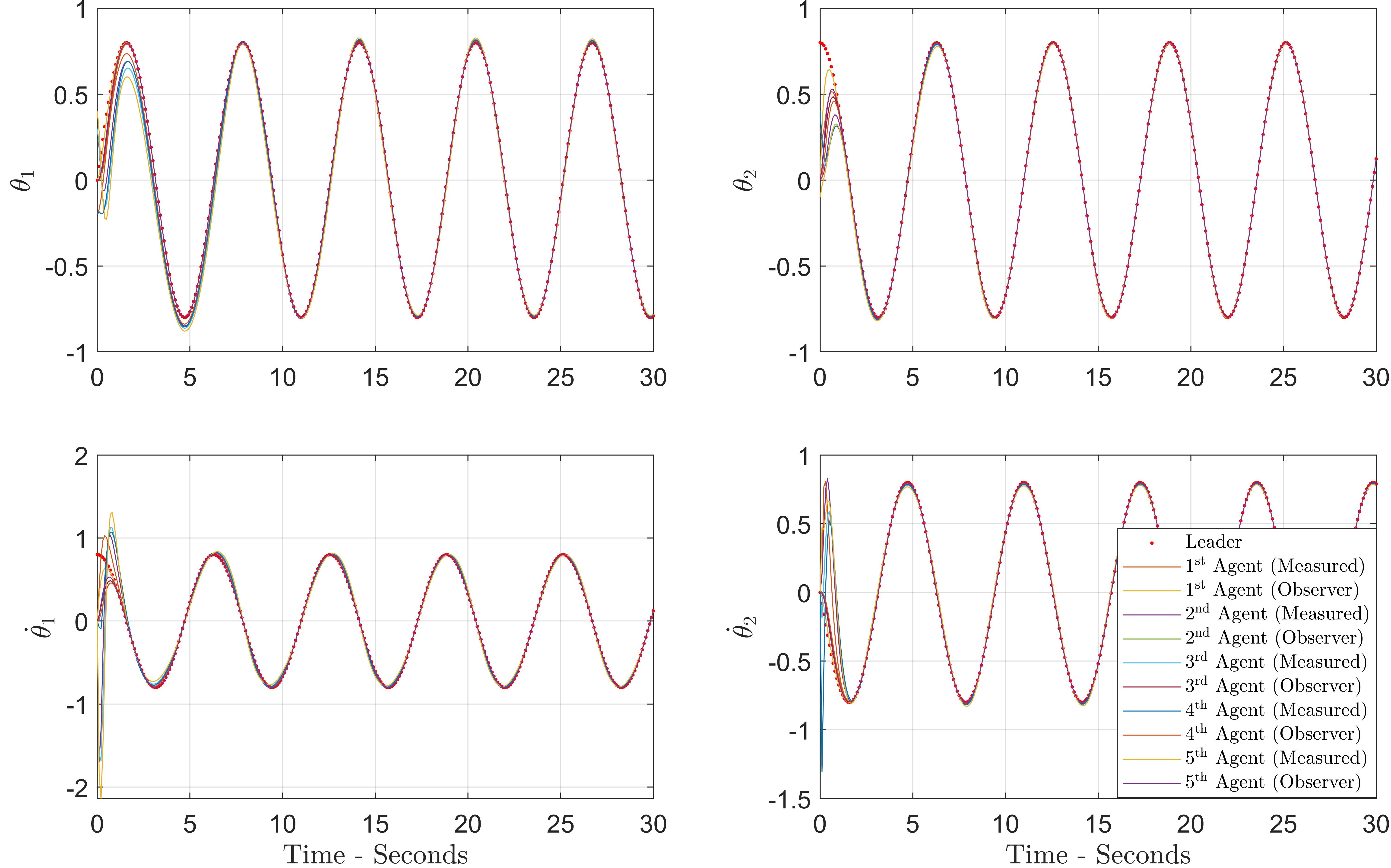}}
\caption{Joint angle tracking response for each agent}
\label{fig4}
\end{figure}
The tracking response of joint angles from the second layer in comparison to the desired signals originating from the first layer and the leader's signal is represented in Fig. \ref{fig4}. It is observed that all signals rapidly converge to align with the leader's signal within the initial seconds.
Fig. \ref{fig5} shows how quickly the Neural Network weights settle for all robotic agents. 
\begin{figure}[!t]
\centerline{\includegraphics[width=\columnwidth]{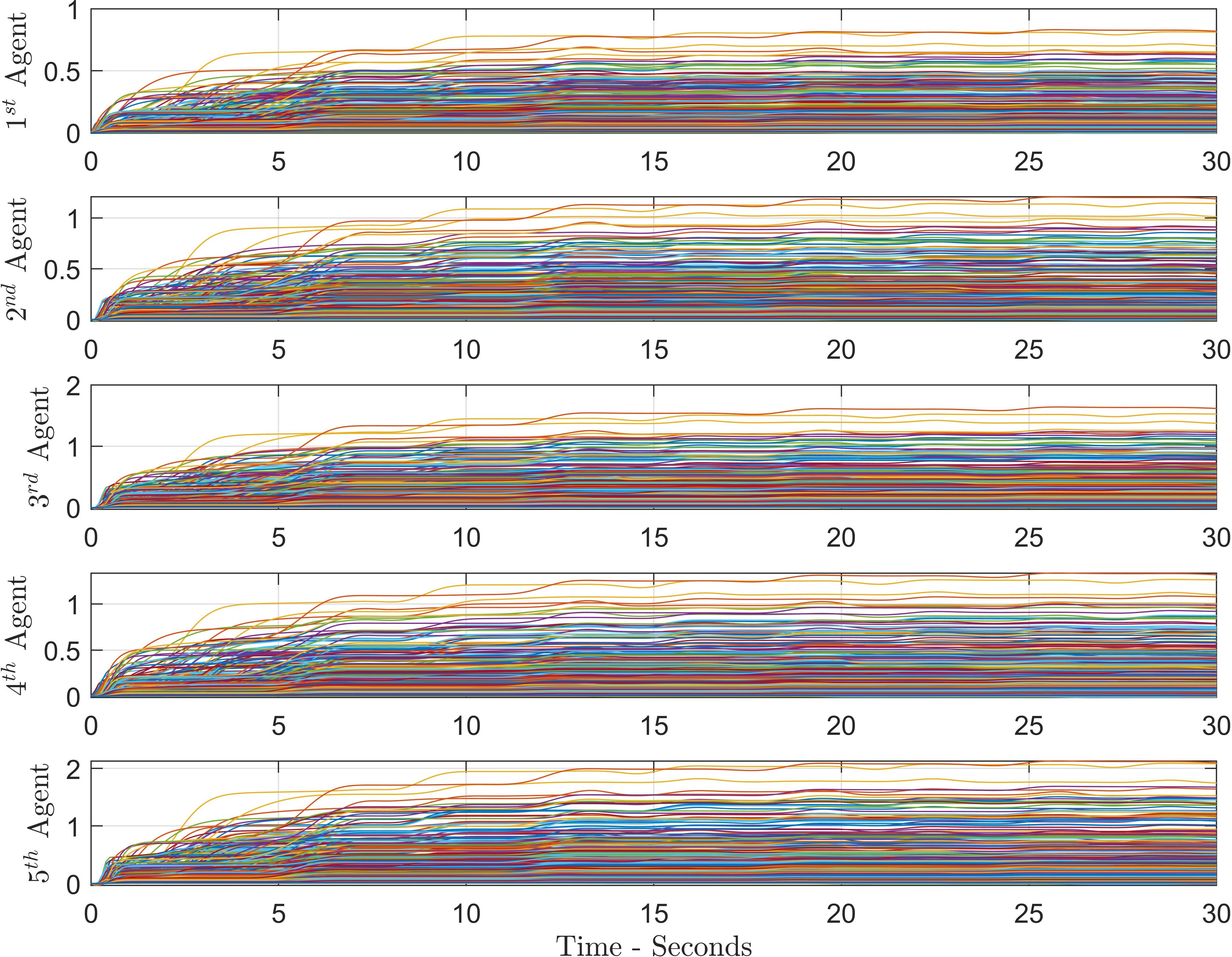}}
\caption{NN weights Convergence for all agents.}
\label{fig5}
\end{figure}
Moreover, we graphed the NN approximation associated with the unknown dynamic variables \scalebox{0.9}{\( H_i(\chi_i) \)} for each agent in Fig.~\ref{fig6}.  
In our controller, all system parameters are treated as having unstructured uncertainty, indicating that the control algorithm is environment-independent. Whether the system operates underwater, where buoyancy forces vary with the depth of a robot's arm, or in space, where the inertia matrix is unknown, the algorithm adapts accordingly. The identified system nonlinear dynamics can be saved and reused in the controller, even after the system restarts.
\begin{figure}[!t]
\centerline{\includegraphics[width=\columnwidth]{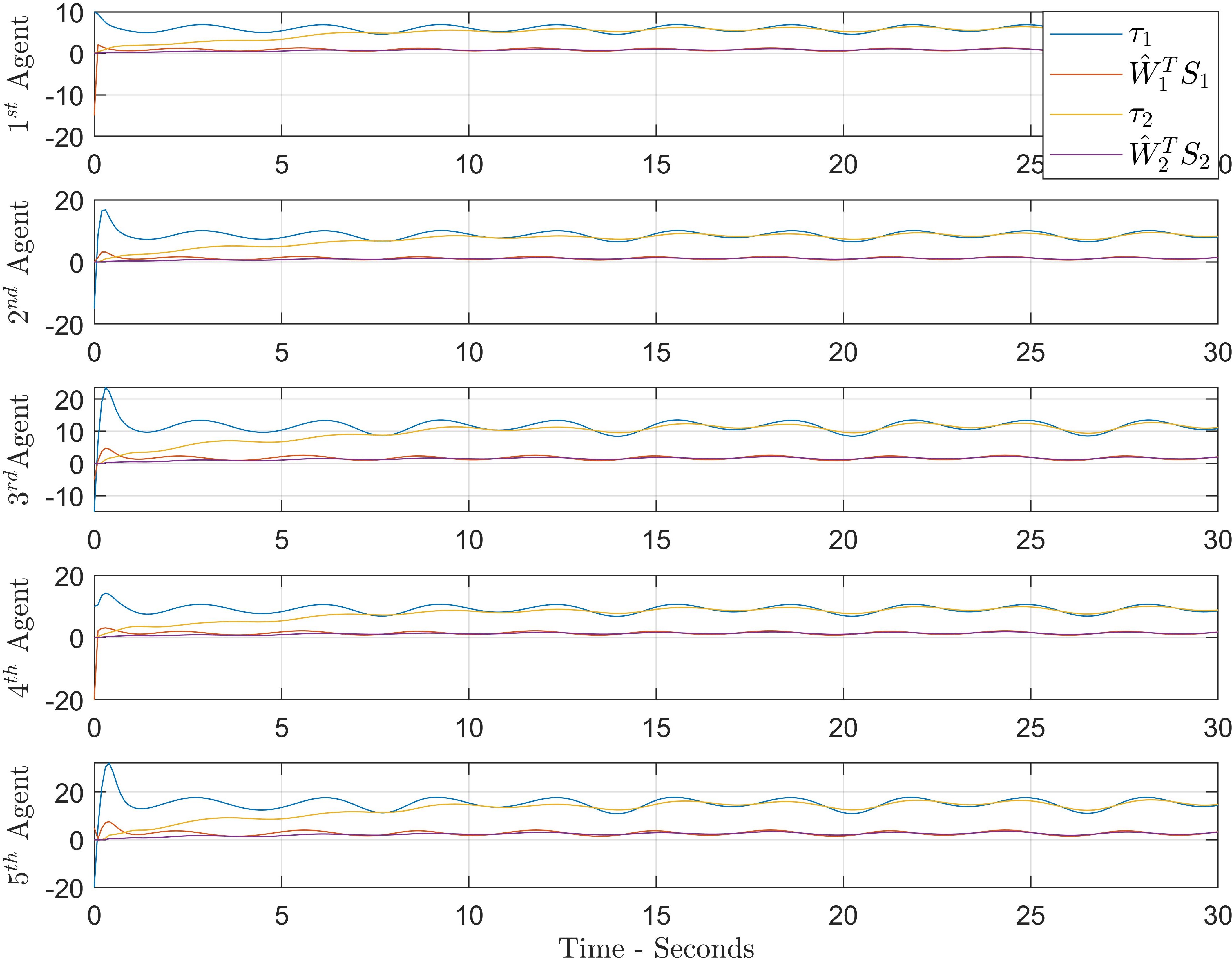}}
\setlength{\belowcaptionskip}{-15pt} 
\caption{NN approximation}
\label{fig6}
\end{figure}
\section{CONCLUSION}
\label{conclusion}
In this study, The challenge of implementing composite synchronization and adaptive learning control for a network of multi-robot manipulators, with complete nonlinear uncertain dynamics was successfully met. The novel approach contains a distributed cooperative estimator in the first layer for estimating the virtual leader's states, and a decentralized adaptive learning controller in the second layer to track the leader's states while identifying each robot's distinct nonlinear uncertain dynamics. The key strengths of This method include simultaneously achieving synchronization control and learning robots' complete nonlinear uncertain dynamics in a decentralized manner. This improves robotics control in space and underwater settings, where system dynamics are typically uncertain. The method's effectiveness is verified through detailed mathematical validation and simulations.
\bibliographystyle{IEEEtran}
\bibliography{root}

\begin{thebibliography}{10}
\providecommand{\url}[1]{#1}
\csname url@rmstyle\endcsname
\providecommand{\newblock}{\relax}
\providecommand{\bibinfo}[2]{#2}
\providecommand\BIBentrySTDinterwordspacing{\spaceskip=0pt\relax}
\providecommand\BIBentryALTinterwordstretchfactor{4}
\providecommand\BIBentryALTinterwordspacing{\spaceskip=\fontdimen2\font plus
\BIBentryALTinterwordstretchfactor\fontdimen3\font minus \fontdimen4\font\relax}
\providecommand\BIBforeignlanguage[2]{{%
\expandafter\ifx\csname l@#1\endcsname\relax
\typeout{** WARNING: IEEEtran.bst: No hyphenation pattern has been}%
\typeout{** loaded for the language `#1'. Using the pattern for}%
\typeout{** the default language instead.}%
\else
\language=\csname l@#1\endcsname
\fi
#2}}

\bibitem{cui2012mutual}
R.~Cui and W.~Yan, ``Mutual synchronization of multiple robot manipulators with unknown dynamics,'' \emph{Journal of Intelligent \& Robotic Systems}, vol.~68, pp. 105--119, 2012.

\bibitem{seenu2020review}
N.~Seenu, K.~C. RM, M.~Ramya, and M.~N. Janardhanan, ``Review on state-of-the-art dynamic task allocation strategies for multiple-robot systems,'' \emph{Industrial Robot: the international journal of robotics research and application}, vol.~47, no.~6, pp. 929--942, 2020.

\bibitem{jandaghi2023motion}
E.~Jandaghi, X.~Chen, and C.~Yuan, ``Motion dynamics modeling and fault detection of a soft trunk robot,'' in \emph{2023 IEEE/ASME International Conference on Advanced Intelligent Mechatronics (AIM)}.\hskip 1em plus 0.5em minus 0.4em\relax IEEE, 2023, pp. 1324--1329.

\bibitem{ghafoori2024novel}
S.~Ghafoori, A.~Rabiee, M.~Jouaneh, and R.~Abiri, ``A novel seamless magnetic-based actuating mechanism for end-effector-based robotic rehabilitation platforms,'' 2024.

\bibitem{yuan2017formation}
C.~Yuan, S.~Licht, and H.~He, ``Formation learning control of multiple autonomous underwater vehicles with heterogeneous nonlinear uncertain dynamics,'' \emph{IEEE transactions on cybernetics}, vol.~48, no.~10, pp. 2920--2934, 2017.

\bibitem{hazon2008redundancy}
N.~Hazon and G.~A. Kaminka, ``On redundancy, efficiency, and robustness in coverage for multiple robots,'' \emph{Robotics and Autonomous Systems}, vol.~56, no.~12, pp. 1102--1114, 2008.

\bibitem{huang2015adaptive}
J.~Huang, C.~Wen, W.~Wang, and Y.-D. Song, ``Adaptive finite-time consensus control of a group of uncertain nonlinear mechanical systems,'' \emph{Automatica}, vol.~51, pp. 292--301, 2015.

\bibitem{abdelatti2018cooperative}
M.~Abdelatti, C.~Yuan, W.~Zeng, and C.~Wang, ``Cooperative deterministic learning control for a group of homogeneous nonlinear uncertain robot manipulators,'' \emph{Science China Information Sciences}, vol.~61, pp. 1--19, 2018.

\bibitem{liu2021adaptive}
Q.~Liu, D.~Li, S.~S. Ge, R.~Ji, Z.~Ouyang, and K.~P. Tee, ``Adaptive bias rbf neural network control for a robotic manipulator,'' \emph{Neurocomputing}, vol. 447, pp. 213--223, 2021.

\bibitem{rodriguez2004mutual}
A.~Rodriguez-Angeles and H.~Nijmeijer, ``Mutual synchronization of robots via estimated state feedback: a cooperative approach,'' \emph{IEEE Transactions on control systems technology}, vol.~12, no.~4, pp. 542--554, 2004.

\bibitem{chung2009cooperative}
S.-J. Chung and J.-J.~E. Slotine, ``Cooperative robot control and concurrent synchronization of lagrangian systems,'' \emph{IEEE transactions on Robotics}, vol.~25, no.~3, pp. 686--700, 2009.

\bibitem{wang2013flocking}
H.~Wang, ``Flocking of networked uncertain euler--lagrange systems on directed graphs,'' \emph{Automatica}, vol.~49, no.~9, pp. 2774--2779, 2013.

\bibitem{dong2019composite}
X.~Dong, C.~Yuan, P.~Stegagno, W.~Zeng, and C.~Wang, ``Composite cooperative synchronization and decentralized learning of multi-robot manipulators with heterogeneous nonlinear uncertain dynamics,'' \emph{Journal of the Franklin Institute}, vol. 356, no.~10, pp. 5049--5072, 2019.

\bibitem{ren2005consensus}
W.~Ren and R.~W. Beard, ``Consensus seeking in multiagent systems under dynamically changing interaction topologies,'' \emph{IEEE Transactions on automatic control}, vol.~50, no.~5, pp. 655--661, 2005.

\bibitem{park1991universal}
J.~Park and I.~W. Sandberg, ``Universal approximation using radial-basis-function networks,'' \emph{Neural computation}, vol.~3, no.~2, pp. 246--257, 1991.

\bibitem{wang2018deterministic}
C.~Wang and D.~J. Hill, \emph{Deterministic learning theory for identification, recognition, and control}.\hskip 1em plus 0.5em minus 0.4em\relax CRC Press, 2018.

\bibitem{wang2009deterministic}
------, \emph{Deterministic learning theory for identification, recognition, and control}.\hskip 1em plus 0.5em minus 0.4em\relax CRC Press, 2009.

\bibitem{yuan2011persistency}
C.~Yuan and C.~Wang, ``Persistency of excitation and performance of deterministic learning,'' \emph{Systems \& control letters}, vol.~60, no.~12, pp. 952--959, 2011.

\end{thebibliography}
\end{document}